\documentclass[aps,twocolumn]{revtex4}
\usepackage{bm}
\usepackage{amsmath}
\usepackage{amssymb}
\usepackage{graphicx}
\usepackage{epsfig}
\usepackage{epstopdf}
\usepackage{mathrsfs}

\begin{document}
\newtheorem{theorem}{Theorem}
\newtheorem{corollary}{Corollary}

\def\be{\begin{equation}}
\def\en#1{\label{#1}\end{equation}}

\newcommand{\rd}{\mathrm{d}}
\newcommand{\vare}{\varepsilon }
  \newcommand{\tb}{\mathbf{t}}
\newcommand{\Phib}{\mathbf{\Phi}}
\newcommand{\Psib}{\mathbf{\Psi}}
  \newcommand{\U}{\mathcal{U}}
 \newcommand{\cH}{\mathcal{H}}

\title{Exact  solution for   a  class  of quantum models  of  interacting    bosons   }

 \author{Valery  Shchesnovich \\
 E-mail: valery@ufabc.edu.br }
 
\affiliation{Centro de Ci\^encias Naturais e Humanas, Universidade Federal do
ABC, Santo Andr\'e,  SP, 09210-170 Brazil }

\begin{abstract}
Quantum models of interacting bosons have a wide range of applications, including the propagation of optical modes in nonlinear media, such as the $k$-photon down-conversion. Many of these models are related to nonlinear deformations of finite group algebras and, in this sense, are exactly solvable. While advanced group-theoretic methods were developed to study the eigenvalue spectrum, in quantum optics, the primary focus is not on the spectrum of the Hamiltonian but rather on the evolution of an initial state  -- such as the generation of optical signal modes by a strong pump mode propagating through a nonlinear medium. I propose a simple and general method to solve the state evolution problem, applicable to a broad class of quantum models of interacting bosons. For the k-photon down-conversion model and its generalizations, the solution to the state evolution problem is expressed as an infinite series expansion in powers of the propagation time, with coefficients determined by a recursion relation involving only a single polynomial function. This polynomial function is unique to each nonlinear model. As an application, I compare the exact solution of the parametric down-conversion process with the semiclassical parametric approximation.
 \end{abstract}
\maketitle

\section{Introduction}
Obtaining an exact solution to a physically relevant model remains highly valuable, even in the era of widespread computer simulations.  Some crucial features  of the solution, such as the  asymptotic  or qualitative behavior,  can  only be studied  by  the analytical approach.   Over time, widely applicable algebraic methods have been developed to derive exactly solvable models and obtain exact solutions to eigenvalue problems. One of the earliest such methods was Bethe’s ansatz for the Heisenberg antiferromagnetic chain model  \cite{BetheAnsatz},     later extended to the Bose gas with a zero-range interaction potential  \cite{BoseGas}  and  the one-dimensional  Hubbard model \cite{1DHubbard}. Other algebraic techniques have since been introduced, such as the factorization method for second-order differential operators, which expresses the quantum Hamiltonian as a product of two first-order ladder operators    \cite{Factoriz},   the  Darboux transformations \cite{Darboux}, and the Quantum   Inverse Scattering Method   \cite{QISM}.  

Exactly solvable models are also prevalent in quantum optics. The effective potential method for generating function approach has been applied to the Dicke model  \cite{DickeModel}  and spin systems  \cite{SpinSystems}. Additionally, the Bethe ansatz has been used to derive the energy spectra of the three-boson model \cite{3BosonBethe}. Furthermore, it has been shown that some quantum Hamiltonians of fourth order in boson creation and annihilation operators admit exact solutions for parts of the energy spectrum due to hidden symmetries  \cite{4BosonModels}. These models, which include the  $k$-photon down-conversion and multiple-photon cascades, belong to the class of quasi-exactly solvable models, where only a subset of the energy spectrum can be analytically determined \cite{QESMod}. Group-theoretic methods sometimes allow mapping the Hamiltonian onto the generators of a deformed Lie algebra, enabling the derivation of eigenvalues and eigenfunctions for nonlinear quantum optical Hamiltonians, such as those describing the second-harmonic generation  \cite{GroupMeth,LieAlg3Boson,GroupMeth2}. These approaches have recently been employed to analyze the energy spectra of a wide range of nonlinear boson models \cite{LieAlgGen1,LieAlgGen2}.

In the rapidly evolving field of exactly solvable models, research has predominantly focused on determining the spectrum of the Hamiltonian and its  associated eigenstates. However, in quantum optics, the primary objective is   to solve the  state evolution   problem rather than   the eigenvalue problem. Importantly, the eigenstates of the  complete  Hamiltonian do not necessarily correspond to the optical energy, which is defined by the free-propagation component of the Hamiltonian. A key example is the cornerstone of quantum optics: the two-photon parametric down-conversion in a nonlinear medium with second-order nonlinearity \cite{Wu86,Slusher87} (see also the reviews  \cite{ReviewSq,Couteau18,Zhang21}). The recent experimental realization of the long-sought three-photon spontaneous parametric down-conversion \cite{3phDC} further expands the class of integrable models relevant to quantum optics.

Analytical approaches, such as deformed Lie algebras and the Quantum Inverse Scattering Method, often involve highly technical mathematical machinery for deriving exact solutions to solvable models. This complexity may explain why  popular   models, including  the $k$-photon down-conversion  \cite{ClasskSq,GenSqueez}, are   analyzed in the physics literature using alternative methods. These include semiclassical approximations, WKB-like techniques \cite{WKB4GenFun,Hillery84, WKBpump, Crouch88}, and numerical simulations \cite{Reid88, Drobny92, Buzek93, Drobny94, Hillery95}  supported by conservation-law-based reductions. Theoretical studies in the physics literature, e.g. Refs.  \cite{BeyondPA1, BeyondPA2}, continue to rely on various approximation schemes in order to go beyond the  quasi-classical  parametric approximation.

The aim of this work is to propose a simple algebraic method for deriving exact solutions to the state evolution problem for a broad class of integrable models of interacting bosons. The method is applied to obtain an exact solution to the problem of optical signal mode(s) generation by a pump  mode propagating in a nonlinear medium. The approach is particularly relevant to quantum optical models, such as  the $k$-photon down-conversion and similar  processes.

In Section \ref{sec2},  I describe the class of quantum models of interacting bosons that can be analyzed   by the proposed unified algebraic framework. Section \ref{sec3}  presents the core methodology, where I develop the unified algebraic framework for solving the state evolution problem. In this section, Theorem 1 and Corollary 1 provide explicit solution to the state evolution problem relevant to quantum optics applications. The derived solution is then verified by direct substitution into the Schr\"odinger equation in the Fock space (Subsection \ref{sec3A}). Subsections \ref{sec3B}, \ref{sec3C}, and \ref{sec3D}  explore and discuss key mathematical features of the solution. Section \ref{sec4} examines some applications, including optical signal mode generation via pump mode propagation in a nonlinear medium. Subsection  \ref{sec4A} discusses the scaling transformation of propagation time for a strong coherent pump mode, while Subsection \ref{sec4B} compares the exact solution with the parametric approximation for squeezed state generation. Finally, Section \ref{sec5} provides a summary of the results and outlines open   questions for future research.

\section{Models of interacting bosons in quantum optics}
\label{sec2}

In quantum optics, models of interacting bosons naturally arise due to the propagation of optical modes in nonlinear media, provided that the phase-matching conditions are satisfied. A key area of interest is the generation of signal modes from the vacuum state through interactions with a strong pump mode. In a lossless medium, conservation of energy — governed by the Manley–Rowe relations  \cite{NonlOpt} — ensures that the  total optical  energy remains constant during propagation. The resulting model Hamiltonian can be decomposed into two parts: the free-propagation term,   $\hat{H}_0$, which represents the  optical     energy and is quadratic in the boson creation and annihilation operators, and the interaction term,  $\hat{H}_{1}$, which is of higher order in the boson operators and governs photon conversion processes in the nonlinear medium. When the phase-matching conditions hold, the interaction term preserves the total optical energy, leading to the commutator condition $[\hat{H}_0,\hat{H}_1]=0$.

The class of interaction models described above includes the  $k$-photon down-conversion processes \cite{GenSqueez}, such as the generation of squeezed states of light via parametric down-conversion in a second-order nonlinear medium (e.g., Refs. \cite{Couteau18,Zhang21}). More recently, three-photon spontaneous parametric down-conversion has been experimentally demonstrated  \cite{3phDC}, further expanding the range of accessible nonlinear optical interactions.

In its simplest form, the  $k$-photon down-conversion process involves the interaction of a single pump mode and a single signal mode. This interaction is governed by the following two-mode Hamiltonian \cite{GenSqueez}:  $\hat{H}=\hat{H}_0 +\hat{H}_1$, 
\be
\hat{H}_0 = \hbar \omega_0\hat{a}^\dag\hat{a} + \hbar \omega \hat{b}^\dag\hat{b}, \quad \hat{H}_1 
= \hbar\Omega\left\{ \hat{a}^\dag \hat{b}^k + \hat{a}(\hat{b}^\dag)^k\right\}
\en{Ham_sq}
where $\hat{a}$ and $\hat{b}$ are the annihilation operators for the pump and signal modes, respectively, and $\hbar\Omega$  characterizes the photon conversion strength specific to the nonlinear medium. The phase-matching condition requires that  $\omega_0 = k\omega$, which ensures the commutator relation $[\hat{H}_0,\hat{H}_1]=0$.  The Hilbert space of the two interacting modes can then be decomposed into a direct sum of invariant subspaces, each labeled by a discrete eigenvalue of  $\hat{H}_0$, given by  $\mathcal{E} = \hbar\omega_0  N+ \hbar \omega \ell$,  where $N$ and $\ell$   are integers satisfying  $\ell \le k-1$. The dimension of each invariant subspace  $\cH_{N,\ell}$ is  given by $\mathrm{dim}\cH_{N,\ell} = N+1$.

In this framework, the initial value problem of interest in quantum optics is the conversion of the optical pump mode ($\hat{a}$)  into the signal mode ($\hat{b}$). Typically, the system is initialized with the signal mode in the vacuum state ($\ell=0$), meaning that all energy initially resides in the pump mode.

For example, in a quadratic nonlinear medium ($k=2$), assuming a strong pump and short propagation times  \cite{WKBpump}, the standard approach treats the pump mode as a coherent state with large amplitude   $\alpha$ \cite{ReviewSq}:
\be
|\alpha\rangle\equiv e^{-|\alpha|^2/2} \sum_{N=0}^\infty \frac{(\alpha\hat{a}^\dag)^N}{N!}|Vac\rangle, 
\en{CohSt}
where  $|Vac\rangle$  denotes the vacuum state of the multimode system, satisfying  $\hat{a}|Vac\rangle = 0$. In the parametric approximation, the boson operator  $\hat{a}$  is replaced by its classical mean-field amplitude,  $\hat{a}\to \alpha$, which converts the interaction Hamiltonian $\hat{H}_1$  in Eq. (\ref{Ham_sq}) (for $k=2$) into its  semiclassical approximation. This approximation reduces the problem to an analytically solvable system governed by a quadratic Hamiltonian in $\hat{b}$ and $ \hat{b}^\dag$, leading to an explicit solution in the form of a squeezed state (for details, see Refs. \cite{Couteau18,Zhang21}).

While the semiclassical approximation has been instrumental in understanding nonlinear optical interactions, recent studies have revisited the two-photon down-conversion process beyond the parametric approximation, incorporating a full quantum treatment through numerical simulations and analytical approximations  \cite{BeyondPA1, BeyondPA2}.  Additionally, for higher-order ($k\ge 3$) processes described by Eq. (\ref{Ham_sq}), the parametric approximation fails dramatically. Specifically, the norm of the quantum state evolved under the parametric approximation diverges in finite time \cite{ClasskSq}, highlighting the need for a fully quantum mechanical treatment that includes pump mode quantum effects \cite{Buzek93,Drobny94}.  

 Remarkably,  optical  model Hamiltonians,  such as the model  in Eq.~(\ref{Ham_sq}), admit an exact  relatively simple solution to the   evolution problem governing the conversion of the pump mode into the signal mode(s). Moreover, this exact solution follows from a unified mathematical structure applicable to a broad class of similar models. These  models share a key feature:    the Hilbert space is partitioned into finite-dimensional invariant subspaces connected by two ladder operators which are Hermitian-conjugate parts of the Hamiltonian.

Consider, for example, the class of models described by Eq. (\ref{Ham_sq}). By introducing the ladder operator for the  $k$-photon down-conversion process, defined as  $\hat{A} \equiv \hat{a}^\dag\hat{b}^k$, the interaction Hamiltonian in Eq.~(\ref{Ham_sq})  can be rewritten in the form:
\be
\hat{H}_1 = \hat{A}+ \hat{A}^\dag.
\en{H_1} 
This ladder-operator formulation is a key prerequisite for applying the algebraic approach introduced in the next section. Another essential feature is the partitioning of the Hilbert space  $\cH$  into invariant subspaces labeled by a full set of commuting conserved quantities with discrete values. In the case of Eq. (\ref{Ham_sq}), these conserved quantities are given by $\mathcal{I} = (N,\ell)$ leading to the Hilbert space decomposition into a direct sum of the invariant subspaces:
\be
\cH = \sum_\mathcal{I}  \cH_\mathcal{I},
\en{GTP1}
where each invariant subspace  $\cH_\mathcal{I}$ has finite dimension.  
 All higher-order boson interaction models that preserve an equivalent of optical energy, as given by the free-propagation term  $\hat{H}_0$, belong to this class. These include exactly integrable boson models whose energy spectra have been analyzed in Refs.  \cite{GroupMeth,GroupMeth2,LieAlg3Boson,LieAlgGen1,LieAlgGen2}. 

The approach introduced in Section \ref{sec3}  applies to a broad class of models, which generalize the model in   Eq.  (\ref{Ham_sq}) and are    characterized by the following ladder operator:
\be 
  \hat{A}   =  (\hat{a}^\dag)^m \prod_{s=1}^S \hat{b}_s^{k_s},
 \en{models}
where $m\ge 1$, $S\ge 1$  and $k_s\ge 1$  are arbitrary integer parameters. Here,   $\hat{a}$   represents the pump mode, while the additional modes labeled by $s$  correspond to signal modes, which are initially in a quantum state annihilated by $ \hat{A}$. The approach developed in the next section also applies  when the  signal modes  are  initially populated  below the optical energy conversion threshold. For example, in the case of  $S=1$  in Eq. (\ref{models}), where the signal mode initially contains  $\ell$  photons, the method remains applicable as long as   $\ell \le k_1-1$.

\section{Solution to the  evolution problem}
\label{sec3}

Similar to the approaches discussed in the Introduction, the method presented below is entirely algebraic. Like the factorization method \cite{Factoriz}, it relies on ladder operators; however, these operators appear as partitions of the quantum Hamiltonian of a many-body system rather than as the factorization of  a single-particle Schrödinger equation. Unlike the advanced group-theoretic methods used in Refs. \cite{GroupMeth,LieAlg3Boson,GroupMeth2,LieAlgGen1,LieAlgGen2},  which rely on Lie algebras and their deformations, the present  approach uses only elementary algebra. Throughout this section, we will work in the interaction picture, where the Hamiltonian is decomposed into two commuting parts, $\hat{H} = \hat{H}_0 + \hat{H}_1$, with   $[\hat{H}_0,\hat{H}_1]=0$.

 As discussed in the previous section, the method is based on two key assumptions:
 
\noindent\textit{(i).} Partitioning of the Hilbert space:
The Hilbert space $\cH$  of the quantum system can be decomposed into a direct sum (which may be infinite) of finite-dimensional invariant subspaces with respect to the interaction Hamiltonian  $\hat{H}_1$. Without loss of generality, we label these invariant subspaces by a discrete integer  $N\ge 0$:
 \be
 \cH = \sum_N \cH_N,\quad \cH_N = \mathrm{Span}\{|\Psi^{(N)}_0\rangle,|\Psi^{(N)}_1\rangle,\ldots,|\Psi^{(N)}_{N}\rangle  \}.
 \en{H_N}
 If multiple subspaces share the same dimension, an additional label can be introduced to distinguish them. Since we primarily focus on a single invariant subspace  $\cH_N$, we will omit the index  $N$  in what follows for simplicity.
 
\noindent\textit{(ii).} Ladder-Operator Structure of the Interaction Hamiltonian:
The interaction Hamiltonian $\hat{H}_1$	  can be written as the sum of two Hermitian-conjugate ladder operators as in Eq. (\ref{H_1}). 
These ladder operators act within the subspace   $\cH_N$ such that only the nearest-neighbor transitions are allowed, i.e.,
\be
 \langle\Psi_n| \hat{A}|\Psi_{n+1}\rangle\ne0, \quad  \langle\Psi_{n+1}| \hat{A}^\dag|\Psi_{n}\rangle\ne0.
 \en{GTP2}
The overall phases of the basis states in  $\cH_N$   can always be chosen so that the  non-zero  matrix elements in Eq.~(\ref{GTP2}) are real and non-negative. We can therefore introduce a set of   numbers $ \beta_n\ge0 $  (which may depend on  $N$) satisfying
 \be 
\hat{A}|\Psi_{n+1}\rangle = \sqrt{\beta_n} |\Psi_{n}\rangle, \quad \hat{A}^\dag|\Psi_n\rangle= \sqrt{\beta_n}|\Psi_{n+1}\rangle.
\en{AAc}
Since  invariant subspace $\cH_N$ has dimension  $N+1$, we must have  $\beta_N = 0$   to ensure proper termination of the ladder structure.

It is important to note that the only model-dependent quantity in this formulation is the set of parameters  $\beta^{(N)}_n$, which uniquely characterize a given boson interaction model. The latter parameter accounts also  for any  additional   features  of the basis in   $\cH_N$ (for instance, when there are more than one subspace of dimension $N+1$, as discussed in Section \ref{sec2},   we have to introduce an additional index to distinguish them, see also Section \ref{sec4} below, for more details).

In the following,   two  more operators will be needed, in addition to the ladder operators defined in Eq.~(\ref{AAc}). The first is the state-number operator $ \hat{n}$  in  each subspace $\mathcal{H}_N$, which satisfies there 
\be
\hat{n}|\Psi_n\rangle = n |\Psi_n\rangle, \quad n=0,\ldots, N.
\en{GTP3}
The second is the commutator of the ladder operators, defined as   
\be
\hat{B}\equiv [\hat{A},\hat{A}^\dag].
\en{GTP4}
From  Eq. (\ref{AAc})  it follows that  $\hat{B}$  is a scalar function of $\hat{n}$:
 \be
 \hat{B}  =  B(\hat{n}), \quad B(n) \equiv  \beta_n - \beta_{n-1},
\en{Comm}
where for $n=0$ we set $\beta_{n-1}\equiv 0$. 
The ladder operators $\hat{A},\hat{A}^\dag$ satisfy the following commutation relations with the state-number operator  
 \be
 [\hat{n},\hat{A}^\dag] = \hat{A}^\dag,\quad [\hat{n},\hat{A}] = -\hat{A}.
 \en{nAAc}
 The relations in Eq. (\ref{nAAc}) resemble those of the standard bosonic creation and annihilation operators. However, in our case,  $\hat{B}$  is not a scalar  (multiplied by the identity operator) but a function of  $\hat{n}$. In this interpretation,  Eq.~(\ref{Comm}) represents a nonlinear deformation of the standard bosonic algebra.

 For any scalar function $F(x)$, it follows   from Eq. (\ref{nAAc}) that 
\be
F(\hat{n}) \hat{A}^\dag = \hat{A}^\dag F(\hat{n}+1).
\en{FA}
Using  Eq. (\ref{FA})  and the definition of $B(n)$ in Eq.  (\ref{Comm}),  we obtain the following commutation   relation:
\begin{eqnarray}
\label{Comm2}
&&\hat{A}(\hat{A}^\dag)^m = (\hat{A}^\dag)^m \hat{A} + \sum_{l=0}^{m-1}(\hat{A}^\dag)^lB(\hat{n})(\hat{A}^\dag)^{m-1-l}\nonumber\\
&&=(\hat{A}^\dag)^m \hat{A} + (\hat{A}^\dag)^{m-1}  \sum_{s=0}^{m-1} B(\hat{n} +s).
\end{eqnarray}

In view of possible applications to the  pump mode conversion to signal modes in nonlinear optical media, we focus on the initial value problem, where the initial state is annihilated by one of the ladder operators
— specifically   $\hat{A}$.
 That is, the initial quantum state is given by 
 \be
 |\psi(0)\rangle = \sum_{N=0}^\infty c_N |\Psi^{(N)}_0\rangle.
 \en{psi0}
At positive times, the quantum state in the interaction picture evolves as $|\psi(\tau)\rangle = e^{-i\tau\hat{H}_1 }|\psi(0)\rangle$,  where we introduce a dimensionless rescaled propagation time  $\tau $. Since we are working within a single subspace $\cH_N$, our goal is to determine the evolution of the state  $|\Psi^{(N)}_0\rangle$. It is convenient to expand this evolved state using the ladder operators, omitting the index  $N$  of the subspace  $\cH_N$  for simplicity:
\be
e^{-i\tau(\hat{A}+\hat{A}^\dag)}|\Psi_0\rangle = \sum_{n=0}^N \gamma_n(\tau) (-i\hat{A}^\dag)^n|\Psi_0\rangle.
\en{psit} 
The factor  $(-i)^n $ is introduced to ensure that the coefficients  $\gamma_n$  remain real.   Using  Eq. (\ref{AAc}),    we can  relate  $\gamma_n$ 
 and  the normalized quantum amplitudes $\psi_n(\tau)$ of the   expansion of the state in Eq. (\ref{psit}) over the basis in Eq. (\ref{H_N})  in $\cH_N$:
\be
\psi_n \equiv \langle \Psi_n|\psi(\tau)\rangle = (-i)^n\gamma_n \sqrt{\prod_{\ell=0}^{n-1}\beta_\ell}. 
\en{amplit}

Solving the state evolution problem in Eq.~(\ref{psit}) requires finding a general expression for the following quantum state
\be
|A_m\rangle\equiv (\hat{A} +\hat{A}^\dag)^m|\Psi_0\rangle
\en{GTP5}
 for an arbitrary $m\ge 0$, which appears  in  the expansion of the evolution operator $e^{-i\tau(\hat{A}+\hat{A}^\dag)}$.  
To find the explicit expression for the state in Eq. (\ref{GTP5}), we analyze   the first few examples, starting from the initial condition  $|A_0\rangle=|\Psi_0\rangle$.   Applying  $\hat{A} +\hat{A}^\dag$ to the state $|A_{m-1}\rangle$, and utilizing 
Eqs.~(\ref{Comm}) and  (\ref{Comm2}), along with the condition  $\hat{A}|\Psi_0\rangle=0$, we obtain:
 \begin{eqnarray*}  
 &&\!\!\!\!   |A_1\rangle  = (\hat{A}^\dag)^1|\Psi_0\rangle,\\
&& \!\!\!\!   |A_2\rangle  = \left\{(\hat{A}^\dag)^2 + (\hat{A}^\dag)^0\beta_0 \right\}\!\! |\Psi_0\rangle,\\
&& \!\!\!\!   |A_3\rangle =  \left\{(\hat{A}^\dag)^3 + (\hat{A}^\dag)^1\sum_{s_1=0}^1\beta_{s_1} \right\}\!\! |\Psi_0\rangle, \\
&& \!\!\!\!   |A_4\rangle =  \left\{(\hat{A}^\dag)^4 +  (\hat{A}^\dag)^2\sum_{s_1=0}^2\beta_{s_1}  + (\hat{A}^\dag)^0\beta_0\sum_{s_2=0}^1\beta_{s_2} \right\}\!\! |\Psi_0\rangle, \\
&& \!\!\!\!  |A_5\rangle =  \left\{(\hat{A}^\dag)^5 +  (\hat{A}^\dag)^3\sum_{s_1=0}^3\beta_{s_1}  + (\hat{A}^\dag)^1\sum_{s_1=0}^1 \beta_{s_1}\sum_{s_2=0}^{s_1+1}\beta_{s_2} \right\}
\!\! |\Psi_0\rangle, \\
&& \!\!\!\!   |A_6\rangle =  \left\{(\hat{A}^\dag)^6 +  (\hat{A}^\dag)^4\sum_{s_1=0}^4\beta_{s_1}  + (\hat{A}^\dag)^2\sum_{s_1=0}^2 \beta_{s_1}\sum_{s_2=0}^{s_1+1}\beta_{s_2} \right.\\
&& \qquad\left. +(\hat{A}^\dag)^0 \beta_{0}\sum_{s_2=0}^{1}\beta_{s_2}\sum_{s_3=0}^{s_2+1}\beta_{s_3} \right\}\!\! |\Psi_0\rangle.
 \end{eqnarray*}
Observe that in the above sequence, the respective powers of the ladder operator in the expansion of the state  $|A_m\rangle$  are given by $(\hat{A}^\dag)^{m-2l}$  with  $0\le l\le [\frac{m}{2}]$, where the bracket  $[\ldots]$  denotes the integer part. The nested summations in the expression for $|A_m\rangle$ arise from two sources:\\
 \noindent{1.}  Multiplication of the lower-power term $(\hat{A}^\dag)^{(m-1)-2l}$ by $\hat{A}^\dag$.\\ 
  \noindent{2.}  Commutation of  $\hat{A}$  with the higher-power term $(\hat{A}^\dag)^{(m-1)-2(l-1)}$ in the expression for $|A_{m-1}\rangle$. \\
 \noindent The above observations suggest the following result.
\begin{theorem}
The quantum state $|A_m\rangle \equiv (\hat{A} +\hat{A}^\dag)^m|\Psi_0\rangle$ is given by 
\be
|A_m\rangle = \left\{ \sum_{l=0}^{[\frac{m}{2}]}(\hat{A}^\dag)^{m-2l} \prod_{j=1}^l \sum_{s_j=0}^{s_{j-1}+1}\beta_{s_j}\right\}\! |\Psi_0\rangle,
\en{m-state}
where   for $l=0$   the empty  product   is  defined as    $1$, while for $l\ge 1$  the   sum over $s_1$  has the upper limit $s_0+1$ with $s_0 \equiv m-2l-1$. 
\end{theorem}
\noindent\textit{Proof.} The theorem can be  proven by induction. The base case $m=1$ is trivial. Assuming the statement holds for $m$, we multiply  the expression for $|A_m\rangle$   in Eq. (\ref{m-state}) by 
$\hat{A}+\hat{A}^\dag$, then   commute $\hat{A}$ with the   powers of $\hat{A}^\dag$, using Eqs. (\ref{Comm}) and (\ref{Comm2})  (separating the term with $l=0$). We obtain:
 \begin{eqnarray}
\label{proof1}
&&|A_{m+1}\rangle = \Biggl\{\!(\hat{A}^\dag)^{m+1} + (\hat{A}^\dag)^{m-1}\beta_{m-1} \nonumber\\
&&  + \sum_{l=1}^{[\frac{m}{2}]}(\hat{A}^\dag)^{m+1-2l} \prod_{j=1}^l \sum_{s_j=0}^{s_{j-1}+1}\beta_{s_j}\nonumber\\
&&  +  \sum_{l=1}^{[\frac{m-1}{2}]}(\hat{A}^\dag)^{m-1-2l} \beta_{m-1-2l}\prod_{j=1}^l \sum_{s_j=0}^{s_{j-1}+1}\beta_{s_j} \Biggr\}\!|\Psi_0\rangle.
\nonumber\\
\end{eqnarray}
Here, the first two terms originate from the product \mbox{$(\hat{A}+\hat{A}^\dag)(\hat{A}^\dag)^m$,} the third term arises from multiplying the sum with  $1\le\l\le  [\frac{m}{2}] $ from the left by $\hat{A}^\dag$,
and the final term results from multiplication on the left by  $\hat{A}$, taking into account  that the remaining power of   $\hat{A}^\dag$ remains non-negative. Consequently, the upper limit in the last sum is 
$l\le [\frac{m-1}{2}]$. In the following, we analyze the cases of even and odd values of   $m$  separately.

For $m=2p+1$ ($p\ge 1$, as the case of $m=1$ is trivial), the upper limit in both summations in  Eq. (\ref{proof1})  is    $l\le p$.  For  $m+1 = 2(p+1)$,  the upper limit  must be $l\le p+1$, as required by Eq. (\ref{m-state}). To establish the induction step, we perform  the following  two 	 operations on the terms in Eq. (\ref{proof1}). \\
\noindent\textit{1.}  Combining terms to obtain the required $ l = 1$  term in Eq. (\ref{m-state}) for  $m+1$:
We combine the second term on the right-hand side of Eq. (\ref{proof1}) with the  $l = 1$  term in the first summation, yielding
\be
(\hat{A}^\dag)^{m-1}\sum_{s_1=0}^{m-1} \beta_{s_1}.
\en{proof2}
Since we must have   $s_0 = (m+1)-2l-1=m-2$, for $l=1$, it follows that $s_1\le s_0+1$, which satisfies the required summation limits in Eq. (\ref{m-state}).\\
\noindent\textit{2.}  Reindexing the last summation to match the required form in Eq. (\ref{m-state}) for  $m+1$:
To combine the remaining terms in the first summation (for  $2 \leq l \leq p$) with the last summation in Eq. (\ref{proof1}), we introduce a new index $l^\prime \equiv l+1$	 in the last summation.
 The new index satisfies  $2\le l^\prime\le p+1 = [\frac{m+1}{2}]$, which is consistent with  Eq. (\ref{m-state}).  The two terms then take the form:
\begin{eqnarray}
\label{proof3}
&&\sum_{l=2}^{p}(\hat{A}^\dag)^{m+1-2l} \prod_{j=1}^l \sum_{s_j=0}^{s_{j-1}+1}\beta_{s_j}\nonumber\\
&&+  \sum_{l^\prime=2}^{p+1}(\hat{A}^\dag)^{m+1-2l^\prime} \beta_{m+1-2l^\prime}\prod_{j=2}^{l^\prime} \sum_{s^\prime_j=0}^{s^\prime_{j-1}+1}\beta_{s^\prime_j}.
\end{eqnarray}
Here, in accordance with the product over $2\le j\le l^\prime$,  we have shifted the indices in $\beta$'s as follows:  $s_j\to s^\prime_{j+1}$. In the second term in Eq. (\ref{proof3})   the summation over $s^\prime_2$  has the upper limit   $s^\prime_1+1 \equiv s_0 +1$, with $s_0  = m-2l-1 = (m+1)-2l^\prime $.  This precisely matches the index in the first  $\beta$-factor of the second term, which is missing in the first term of Eq. (\ref{proof3}). Consequently, by combining the two sums, the sum over  $s_1$  in the first term is extended to satisfy  $s_1\le (m+1)-2l$, as required by Eq. (\ref{m-state}) for  $m+1$.
For   $l=p+1$ the corresponding sum in Eq. (\ref{m-state}) contains only $\beta_{m+1-2l}=\beta_0$, which arises from the second term in  Eq. (\ref{proof3}) for $l^\prime =p+1$.  Therefore, the two summations in Eq. (\ref{proof3}) combine to reproduce the required sum over	 $2\le l\le [\frac{m+1}{2}]$, confirming that Eq. (\ref{m-state}) holds for  $m+1$. 
 
Now consider the case of an even $m=2p$. In this case, the summation limit in Eq. (\ref{m-state}) is  $1\le l\le p$  for  both $m$ and for $m+1$.  We proceed similarly to the previously analyzed case of  $m=2p+1$. \\
\noindent\textit{1.}  First Step:
The first step is identical to the case of  $m = 2p+1$, leading to  the same result as in Eq. (\ref{proof2}).  By similar arguments, this result coincides with the  $l = 1$  term required by Eq. (\ref{m-state}) for  $m+1 $. \\\noindent\textit{2.}   Second Step:  We now perform the same operations as in the case of $m=2p+1$,  to combine the last summation in Eq. (\ref{proof1}) with the remaining terms in the first summation  (for  $2\le l\le p$). However, since  $[\frac{m-1}{2}] = p-1$, the upper limits in the two summations (after reindexing with    $l^\prime = l+1$) coincide:
\begin{eqnarray}
\label{proof4}
&&\sum_{l=2}^{p}(\hat{A}^\dag)^{m+1-2l} \prod_{j=1}^l \sum_{s_j=0}^{s_{j-1}+1}\beta_{s_j}\nonumber\\
&&+  \sum_{l^\prime=2}^{p}(\hat{A}^\dag)^{m+1-2l^\prime} \beta_{m+1-2l^\prime}\prod_{j=2}^{l^\prime} \sum_{s^\prime_j=0}^{s^\prime_{j-1}+1}\beta_{s^\prime_j}.
\end{eqnarray}
To complete the proof, we must verify that the upper limit for the summation over  $s_1$  in both terms in Eq. (\ref{proof4}) aligns correctly.   Specifically, in the second term, the only contribution to this summation is the first (leftmost)  $\beta$-factor with index $m+1-2l^\prime$. This precisely corresponds to the highest index appearing in the sum over  $s_1$ in the first term, ensuring that we maintain the required bound  $s_1\le m+1-2l$ as dictated by Eq. (\ref{m-state}).
By following similar arguments as in the case of odd  $m$,  we confirm that the upper limit for  $s^\prime_2$  in the second term is aligned with the index of the first  $\beta$-factor, again satisfying Eq. (\ref{m-state}).
 This concludes the proof of the Theorem. Q.E.D. 

The  $m$th term in the power series expansion of the evolution operator, as given in Theorem 1, allows us to express the coefficients  $\gamma_n(\tau)$ in Eq. (\ref{psit}) as some infinite  power series in  $\tau$. To simplify the notations below, we introduce a concise representation for the nested summations of the  $\beta$-factors that appear on the right-hand side of Eq. (\ref{m-state}).
We define  $g^{(0)}_n\equiv 1$  and, for  $l\ge 1$  and  $0\le n\le N$, we set
\be
g^{(l)}_n \equiv  \sum_{s_1=0}^{n}\beta_{s_1}\sum_{s_2=0}^{s_1+1}\beta_{s_2}\ldots \sum_{s_l=0}^{s_{l-1}+1}\beta_{s_l}.
\en{glm}
Both the factor $\beta_n$, defined in Eq. (\ref{AAc}), and consequently  $g^{(l)}_n$, generally vary across different invariant subspaces  $\cH_N$.  For instance, in the applications below, they depend explicitly on the index  $N$. An interesting property of  $g^{(l)}_n$  follows from the condition that the dimension of each invariant subspace satisfies  $\mathrm{dim}(\cH_N)=N+1$.  As discussed at the beginning of this section, this condition imposes $ \beta_N = 0$, which in turn leads to the relation $g^{(l)}_N = g^{(l)}_{N-1}$ in each invariant subspace $\cH_N$. Additionally,  $g^{(l)}_n$ satisfies the recursive relation
\be
g^{(l)}_n = \sum_{s=0}^n\beta_s g^{(l-1)}_{s+1}, \quad g^{(0)}_n=1.
 \en{Recur} 
With these notations, we now state the following corollary to Theorem 1.
\begin{corollary}
The evolution of the initial state given in Eq. (\ref{psi0}), projected onto the invariant subspace $\cH_N$, is given by (omitting the superscript  $N$, for simplicity)
\begin{eqnarray}
\label{psi-state}
&&e^{-i\tau(\hat{A} +\hat{A}^\dag)} |\Psi_0\rangle= \sum_{n=0}^N \gamma_n(\tau) (-i\hat{A}^\dag)^n|\Psi_0\rangle,\nonumber\\
&& \gamma_n(\tau) = \sum_{l=0}^\infty \frac{(-1)^l \tau^{n+2l}}{(n+ 2l)!} g^{(l)}_n.
\end{eqnarray}
In other words,  the amplitude  $\gamma_n(\tau)$ admits  a power series expansion in $\tau$, where the  coefficients are recursively defined  by Eq. (\ref{Recur}). 
\end{corollary}
\noindent\textit{Proof.}  
To prove Eq. (\ref{psi-state}), we substitute the result of Theorem 1 into the power series expansion of the evolution operator and interchange the order of summation. We then introduce a new summation index,  
 $n\equiv m-2l$, treating even and odd values separately, since even (odd) values of  $n$  correspond to even (odd) values of  $m$:
\begin{eqnarray*}
\label{proofC}
&& e^{-i\tau(\hat{A} +\hat{A}^\dag)} |\Psi_0\rangle = \sum_{m=0}^\infty \frac{(-i\tau)^m}{m!}\sum_{l=0}^{[\frac{m}{2}]} g^{(l)}_{m-2l}(\hat{A}^\dag)^{m-2l}|\Psi_0\rangle\\
&& = \sum_{n=0}^N \left(\sum_{l=0}^\infty\frac{(-1)^l\tau^{n+2l}}{(n+2l)!}g^{(l)}_n\right)(-i\hat{A}^\dag)^n|\Psi_0\rangle.
\end{eqnarray*}
The summation inside the parentheses precisely matches the expression for  $\gamma_n(\tau)$  given in Eq. (\ref{psi-state}). Q.E.D.


\medskip
\subsection{Direct verification by substitution}
\label{sec3A}

The validity of the solution can be easily verified by direct substitution into the evolution equation for the amplitudes  $\gamma_n(\tau)$.   Expanding the    Schr\"odinger equation  in the interaction picture for   $|\Psi(\tau)\rangle \equiv e^{-i\tau(\hat{A}+\hat{A}^\dag)}|\Psi_0\rangle$ in the basis $(-i \hat{A}^\dag)^n|\Psi_0\rangle$, and using the definition of $\gamma_n(\tau)$ from  Eq. (\ref{psit}) along with the identity in Eq. (\ref{Comm2}), we obtain:
\be
 \frac{\rd \gamma_n}{\rd \tau } = \gamma_{n-1} - \beta_n \gamma_{n+1}, \quad \gamma_{-1}\equiv \gamma_{N+1}\equiv 0.
\en{gamn-Eq} 
To verify that the amplitudes $\gamma_n(\tau)$  from Eq. (\ref{psi-state}) satisfy Eq. (\ref{gamn-Eq}), we use the following recurrence relation for the coefficients  $g^{(l)}_n$ for  all $l\ge 1$:
 \be
g^{(l)}_n =  g^{(l)}_{n-1}  + \beta_n g^{(l-1)}_{n+1},
 \en{Recur2} 
which follows directly from Eq. (\ref{Recur}).  Differentiating  $\gamma_n $ from  Eq. (\ref{psi-state}) for $n\ge 1$ and using Eq. (\ref{Recur2}),  we obtain (separating the term with $l=0$):
\begin{eqnarray*}
&& \frac{\rd \gamma_n}{\rd \tau } = \frac{\tau^{n-1}}{(n-1)!} + \sum_{l=1}^\infty \frac{(-1)^l\tau^{n-1+2l}}{(n-1+2l)!}g^{(l)}_{n-1}\\
&& + \beta_n  \sum_{l=1}^\infty \frac{(-1)^l\tau^{n-1+2l}}{(n-1+2l)!}g^{(l-1)}_{n+1}\\
&&=\sum_{l=0}^\infty \frac{(-1)^l\tau^{n-1+2l}}{(n-1+2l)!}g^{(l)}_{n-1} - \beta_n\sum_{l^\prime=0}^\infty \frac{(-1)^{l^\prime}\tau^{n+1+2l^\prime}}{(n+1+2l^\prime)!}g^{(l^\prime)}_{n+1}\\
&& = \gamma_n -\beta_n \gamma_{n+1},
\end{eqnarray*}
where we have used that  $\quad  g^{(0)}_n \equiv 1$ and changed the summation index $l^\prime = l-1$ in the last sum. For $n=0$, we proceed similarly using Eq. (\ref{Recur2}):
\begin{eqnarray*}
&& \frac{\rd \gamma_0}{\rd \tau } =  \beta_0\sum_{l=1}^\infty \frac{(-1)^l\tau^{2l-1}}{(2l-1)!}g^{(l-1)}_{1} \\
&& =  -\beta_0\sum_{l^\prime=0}^\infty \frac{(-1)^{l^\prime}\tau^{2l^\prime+1}}{(2l^\prime+1)!}g^{(l^\prime)}_{1} = - \beta_0\gamma_1.
\end{eqnarray*}

Finally, instead of using the algebraic approach developed above, one could alternatively solve Eq. (\ref{gamn-Eq}) by assuming a power series expansion in  $\tau$, where the two-dimensional linear recurrence relation determines the coefficient of the   amplitude 
 $\gamma_n$  at each power of  $\tau$.

\medskip
\subsection{Amplitude  $\gamma_n(\tau)$  is  a holomorphic function of $\tau$}
\label{sec3B}

The solution in Eq. (\ref{psi-state}) remains valid for all finite propagation times $\tau$,  meaning that the infinite power series defining $\gamma_n(\tau)$  is convergent. 
Indeed,  Eq. (\ref{amplit}) and the state normalization condition impose the following  bound
\be
|\gamma_n(\tau)|\le \left({\prod_{\ell=0}^{n-1}\beta_\ell}\right)^{-\frac12},
\en{norm_gam} 
which ensures that the power series in Eq. (\ref{psi-state}) must converge regardless of the values of the $\beta$-factors.  The convergence also  follows from the evident bound:
\be
g^{(l)}_n\le\left(  \sum_{s=0}^{N-1}\beta_s\right)^l=(g^{(1)}_{N-1})^l,
\en{GTP6}
where we use  that  $\beta_n>0$, for $0\le n\le N-1$, and    $\beta_N=0$. 
The  convergence of the power series defining $\gamma_n(\tau)$ follows because it  is bounded by a uniformly convergent series with an infinite radius of convergence:
\be
|\gamma_n(\tau)|\le \sum_{l=0}^\infty \frac{|\tau|^{n+2l}}{(n+2l)!} \left(  \sum_{s=0}^{N-1}\beta_s\right)^l<\infty, \quad \forall |\tau|<\infty, 
\en{Bound_gam} 
Therefore,  $\gamma_n(z)$ is a holomorphic function in the complex plane  $z\in \mathbb{C}$ and   remains bounded on  the real line $\tau= \Re(z)$.  

Next, we estimate the number of terms  $\bar{l}(\epsilon)$  that must be retained in the power series expansion to approximate the exact quantum amplitudes $\gamma_n(\tau)$ within a given error $\epsilon\ll 1$. Setting $\left|\frac{T_{l+1}}{T_l}\right| = \epsilon$, where $T_l$ and $T_{l+1}$ are two consecutive terms in the power series in Eq. (\ref{Bound_gam}), we obtain an estimate for the index of the first discarded term in the series for $\gamma_n(t)$.  The maximum value is attained at  $n=0$, for which case 
\be
\bar{l} (\epsilon) \lesssim  \frac{\tau}{\sqrt{\epsilon}}\sqrt{\sum\limits_{s=0}^{N-1}\beta_s}. 
\en{apprL}

\subsection{Explicit solution: the  beam-splitter example}
\label{sec3C}

Holomorphic functions that remain bounded on the real line include most well-known elementary and special functions. Consequently, in certain special cases, the quantum amplitude  $\gamma_n(\tau)$  must be expressible in terms of compositions of known holomorphic functions, which are also bounded on the real line. However, identifying such a representation, when it exists, is a nontrivial problem.

The beam-splitter example illustrates this point. Consider the quadratic Hamiltonian describing a unitary linear four-port interferometer (commonly known as the beam splitter), which corresponds to Eq. (\ref{Ham_sq}) 
for $k=1$. This case admits an explicit solution in terms of powers of trigonometric functions, which can be derived either by direct integration (in the Heisenberg picture) or by group-theoretic methods. Observing that the interaction Hamiltonian in Eq. (\ref{Ham_sq}) for  $k=1$  satisfies the following commutation relations (where we set  $\hbar= \Omega=1$  for convenience):
\be
[\hat{H}_1,  \hat{a}^\dag] =   \hat{b}^\dag, \quad [\hat{H}_1,  \hat{b}^\dag] =   \hat{a}^\dag,
\en{Comm1}
we obtain the evolved creation operator $\hat{a}^\dag$    (in the Heisenberg picture) as follows 
\be
e^{-i\tau\hat{H}_1} \hat{a}^\dag e^{i\tau\hat{H}_1} = \hat{a}^\dag \cos \tau  - i \hat{b}^\dag \sin \tau. 
\en{Evol_a} 
In each subspace $\cH_N$, the basis introduced in Section \ref{sec2} corresponds to the Fock states, which in this case are given by
\be
 |N-n,n\rangle \equiv  \frac{(\hat{a}^\dag)^{N-n}(\hat{b}^\dag)^n}{\sqrt{(N-n)! n!}}|0,0\rangle ,
\en{GTP7}
where $|0,0\rangle = |Vac\rangle$. In the nomenclature of Section \ref{sec2}, the state  $|\Psi^{(N)}_0\rangle$, which is annihilated by the ladder operator $\hat{A}\equiv   \hat{a}^\dag\hat{b}$, corresponds to the Fock state  $|N,0\rangle$.
In the interaction picture,   Eq. (\ref{Evol_a}) leads to 
\begin{eqnarray}
 \label{Evol_k1}
&& e^{-i\tau\hat{H}_1} |N,0\rangle =\frac{1}{\sqrt{N!}}\left[\hat{a}^\dag \cos \tau  -i\hat{b}^\dag \sin \tau \right]^N|0,0\rangle\nonumber\\
&&= \cos^N \tau \sum_{n=0}^N \binom{N}{n}^\frac12 \left[-i \tan \tau\right]^n|N-n,n\rangle.
\end{eqnarray}
The coefficients  $\gamma_n$  appearing in the equivalent expansion of the state in Eq. (\ref{Evol_k1}), as given in Eq. (\ref{psit}) of Section \ref{sec2}, can also be obtained directly from Eq. (\ref{Evol_k1}), which determines the respective quantum amplitudes as in Eq.~(\ref{amplit}). Observing that the corresponding parameter  $\beta_\ell = (N-\ell)(\ell+1)$ leads to
\[
\prod_{\ell=0}^{n-1}\beta_\ell = (n!)^2\binom{N}{n},
\]
 we obtain the following identity for the rescaled coefficients $n!\gamma_n$:
\be
 \tau^n \sum_{p=0}^\infty \frac{(-\tau^2)^p}{(n+1)\ldots (n+2p)} g^{(p)}_n  = \cos^{N-n}  \tau\sin^n \tau, 
\en{id_k1} 
for all $n=0,1,2,\ldots, N$.  The validity of Eq. (\ref{id_k1}) can be verified by performing a Taylor expansion at  $\tau = 0$  of the function on the right-hand side.

A natural question is whether one can proceed in the inverse direction of the derivation leading to Eq. (\ref{id_k1}), namely, to determine whether the power series expansion on the right-hand side of Eq. (\ref{psi-state}) corresponds to a known combination of elementary and/or special functions.   This appears to be a challenging power series factorization problem,   due to the recursive nature of the coefficients  $g^{(l)}_n$  in Eq. (\ref{Recur}).

\subsection{Matrix form of the power series  }
\label{sec3D}

In the absence of a systematic method to express the infinite power series in Eq. (\ref{psi-state}) as a combination of elementary and special functions, one can compute the expansion coefficients numerically using their recursive definition in Eq. (\ref{Recur}). Notably, this computation can be performed in parallel for all  $0\le n\le N$   by employing a matrix formulation of the recursion for the coefficient vector  
$\mathbf{g}^{(l)}  \equiv (g^{(l)}_0,\ldots,g^{(l)}_N)^T$, where  ``$T$"  denotes transposition.
Indeed, the recursion in Eq. (\ref{Recur}) can be rewritten in matrix form using an  $(N+1)$-dimensional matrix  $\mathbf{B}$, which belongs to a special class of matrices. Denoting  the  $(N+1)$-dimensional  vector of ones, which coincides with the values of $g^{(0)}_n$,  by $\mathbf{1} \equiv (1,\ldots, 1)^T$, we obtain for  $p\ge 1$:
\be
\mathbf{g}^{(p)} = \mathbf{B}^p\cdot\mathbf{1}, \quad  \mathbf{B}_{nl} \equiv \left\{ \begin{array}{cc} \beta_{l-1}, & 0\le l \le \mathrm{min}(n+1,N) \\
0, & n+2\le l\le N. \end{array} \right.
\en{MatB}
Observe that the   matrix $\mathbf{B}$ is almost  lower-triangular, except for the   first nonzero diagonal above the main diagonal. This structure places  $\mathbf{B}$ within the class of the lower Hessenberg matrices.

The key properties of the matrix  $\mathbf{B}$  can be most easily analyzed by considering an explicit example. Setting   $N=4$, we obtain the following   matrix:
\be
\mathbf{B} = \left( \begin{array}{ccccc} 
0 & \beta_0 & 0 & 0& 0  \\
0 & \beta_0 & \beta_1 & 0& 0  \\
0 & \beta_0 & \beta_1 & \beta_2& 0  \\
0 & \beta_0 & \beta_1 & \beta_2 & \beta_3  \\
0 & \beta_0 & \beta_1 & \beta_2& \beta_3  
\end{array}\right).
\en{MatB4}
One immediately observes that the last two rows are identical, a consequence of the definition of  $g^{(l)}_n$  in Eq. (\ref{glm}) and the fact that $\beta_N =0$. As a result, the matrix  $\mathbf{B} $ has rank  
$\mathrm{rank}(\mathbf{B}) = N$. Furthermore,  $\mathbf{B}$  admits a simple $LU$-decomposition. For the above example with $ N = 4$, we obtain:
\be
\mathbf{B} = \mathbf{L} \mathbf{U} \equiv \left( \begin{array}{ccccc} 
1 & 0 & 0 & 0& 0  \\
1 & 1 & 0 & 0& 0  \\
1 & 1 & 1 & 0 & 0  \\
1 & 1 &1 &1 &0 \\
1 &1 & 1 &1& 1 
\end{array}\right) \left( \begin{array}{ccccc} 
0 & \beta_0 & 0 & 0& 0  \\
0 & 0 & \beta_1 & 0& 0  \\
0 & 0 &0 & \beta_2& 0  \\
0 &  0 & 0 & 0 & \beta_3  \\
0 &  0 &  0 &0 & 0  
\end{array}\right).
\en{MatB4LU}
Using this structure, the infinite power series in Eq. (\ref{psi-state}), defining $\gamma_n(\tau)$,   can be rewritten as a series in powers of  $\mathbf{B}$:
\be
{\gamma}_n(\tau)  = \frac{\tau^n}{n!} \sum_{p=0}^\infty \frac{(-\tau^2)^p}{(n+1)\ldots (n+2p)}\left(\mathbf{B}^p\cdot \mathbf{1}\right)_n.
\en{gam_MatB}
Eq.~(\ref{gam_MatB}) provides an efficient framework for numerical computation of the power series coefficients using fast algorithms for computing powers of Hessenberg matrices \cite{Hessen}.

\section{Applications in   quantum optics}
\label{sec4}

To apply the theory developed in Section \ref{sec3}, one must partition the interaction Hamiltonian into a sum of a ladder operator and its Hermitian conjugate, $\hat{H}_1= \hat{A}+\hat{A}^\dag$. Additionally, within the basis states of each invariant subspace of the Hilbert space, one must identify the state $|\Psi_0\rangle$  that is annihilated by the ladder operator: $\hat{A}|\Psi_0\rangle = 0$.

Now, consider the application of this framework to the models described by Eq. (\ref{models}) in Section \ref{sec2}. In the simplest two-mode case ($S=1$) with  $\hat{A} = (\hat{a}^\dag)^m\hat{b}^k$, the invariant subspaces are labeled by the composite index  $\mathcal{I} \equiv  (M,\ell)$, where  $M\ge 0$ and $0\le \ell \le k-1$. In each subspace  $\cH_\mathcal{I}$, the basis states are given by the Fock states of the two modes:
\begin{eqnarray}
\label{model_1}
&& |\Psi^{(\mathcal{I})}_n\rangle \equiv |M-mn,kn+\ell \rangle, \quad  0\le n\le N\equiv \left[ \frac{M}{m}\right] ,\nonumber\\
&&  |M-mn,kn+\ell \rangle \equiv \frac{(\hat{a}^\dag)^{M-mn}(\hat{b}^\dag)^{kn+\ell}}{\sqrt{(M-mn)!(kn+\ell)!}}|Vac\rangle,\nonumber\\
\end{eqnarray} 
where $[\ldots]$ is the integer part. The corresponding parameter $\beta^{(\mathcal{I})}_n$ in each invariant subspace $\cH_\mathcal{I}$ can be determined  from Eqs. (\ref{AAc}) and (\ref{model_1}):
\be
\beta^{(\mathcal{I})}_n = \left[\prod_{i=0}^{m-1}(M-mn-i) \right]\prod_{j=1}^k(kn+\ell+j).
\en{beta_1}
Since $N$   defined in Eq. (\ref{model_1}) must have integer values,  there always exists an integer  $q$  such that  $M = Nm + q$ and   $0 \leq q \leq m-1$. Consequently, as dictated by the finite dimension of $\cH_\mathcal{I}$, we obtain  $\beta^{(\mathcal{I})}_N =0$  in  Eq. (\ref{beta_1}), since at least one factor in the first product vanishes.

In the general multi-mode case of Eq. (\ref{models}),  the invariant subspaces are labeled by the composite index  $\mathcal{J} \equiv  (M,\ell_1,\ldots,\ell_S)$, where  $M\ge 0$ and $0\le \ell_s \le k_s-1$. The corresponding basis states are Fock states, analogous to Eq. (\ref{model_1}). In this case, the  $\beta$-parameter is given by:
\be
\beta^{(\mathcal{J})}_n = \left[\prod_{i=0}^{m-1}(M-mn-i) \right]\prod_{s=1}^S\prod_{j=1}^{k_s}(k_sn+\ell_s+j).
\en{beta_2}

\subsection{Rescaling   propagation   time  by   subspace size  }
\label{sec4A}

 For the class of models in Eq. (\ref{models}), the (dimensionless) propagation time  $\tau$  and the size of the respective Hilbert subspace $\cH_N$  can be combined via a scaling transformation. Consider the simplest two-mode case with the  $\beta$-parameter from Eq. (\ref{beta_1}).  We introduce a rescaled  $\bar{\beta}$-parameter by factoring out the dominant term $M^m \sim (mN)^m$,
\be
 \bar{\beta}_n \equiv \left[\prod_{i=0}^{m-1}\left(1-\frac{mn-i}{M}\right) \right]\prod_{j=1}^k\left(kn+\ell+j\right).
\en{beta_res}
 Additionally, the propagation time is rescaled in each invariant subspace $\cH_N$ as 
 \[
  \bar{\tau}^{(N)} \equiv \sqrt{M^m}\tau \sim \sqrt{(mN)^m}\tau.
  \]
  This rescaling leaves the quantum amplitudes  $\psi_n^{(N)}$  in  Eq.~(\ref{amplit})   invariant. Specifically, the function  $\psi_n^{(N)}(\bar{\tau}^{(N)})$  has  the same power series expansion in the rescaled time  $\bar{\tau}^{(N)}$, with coefficients given by the rescaled  $\bar{\beta}$-parameter in Eq. (\ref{beta_res}). 
 
The above scaling transformation is particularly useful in quantum optical applications where the pump mode corresponds to the output of a strong laser. The expansion of a strong semiclassical pump state, such as the coherent state from Eq. (\ref{CohSt}) with $\alpha\gg 1$, can be truncated by retaining only the Fock states within a relatively small interval around the mean,
\be
\left| \frac{N}{\langle N\rangle} -1\right| = \mathcal{O}\left(\alpha^{-1}\right).
\en{ToREF}
where  $\langle N\rangle = \alpha^2$. The reason for this truncation is that the Poisson distribution of the subspace weights,
\[
\mathcal{P}_N(\alpha) = e^{-\alpha^2}\frac{ \alpha^{2N}}{N!}
\]
 is negligible outside this interval. Consider now the invariant subspaces $\cH_N$ that contain a significant part of the state norm. In all of these invariant subspaces, the scaling of the propagation time is approximately the same, up to a relative error $\mathcal{O}\left(\alpha^{-1}\right)$.

 To illustrate the above point, consider the ``generalized squeezing'' process introduced in Ref. \cite{GenSqueez}, which corresponds to the class of models in Eq. (\ref{models}) with  $S = m = 1$  and an arbitrary  
$k \geq 2$. In this case, the rescaled time is given by
\[
\bar{\tau} = \sqrt{N}\tau.
\]
The maximum propagation time within the parametric approximation for such models was found in Ref. \cite{Buzek93} to be
\[
\tau_c \sim \frac{1}{\alpha}
\]
when the pump mode is in a strong coherent state with $\alpha =\sqrt{\langle N\rangle}$. This translates into our rescaled time as
\[
\bar{\tau}_c\sim 1
\]
 for the significant Fock state components of the coherent state.
\medskip
\subsection{Comparing  the exact solution with the Gaussian squeezed state}
\label{sec4B}

Consider now the most important case of second-order nonlinearity   as described in Eq. (\ref{Ham_sq}) of Section~\ref{sec2} for \mbox{$k=2$.} This case models the spontaneous down-conversion process \cite{Wu86, Slusher87} (see also the reviews \cite{ReviewSq, Couteau18, Zhang21}).

First, let us briefly recall the standard parametric approximation for a strong pump, specifically for the coherent-state pump in Eq. (\ref{CohSt}) of Section \ref{sec2}, where $\alpha \gg 1$. In this approximation, it is assumed that the strong pump remains unchanged during propagation. This allows us to replace the boson operators of the pump mode with a scalar parameter: $\hat{a}\to \alpha$ and $\hat{a}^\dag \to \alpha$. which gives the approximation its name. The approximate Hamiltonian can be derived from Eq. (\ref{Ham_sq}) using this procedure:
\be
\hat{H}^{(\alpha)}_1 \equiv  \hat{H}_1\left[{\begin{array}{c}\hat{a}\to \alpha \\ \hat{a}^\dag\to \alpha \end{array}}\right]= \hbar\Omega \alpha(\hat{b}^2 +\hat{b}^\dag{}^2).
\en{H_par}
The above  Hamiltonian can be   mapped onto the generators of the $SU(1,1)$-group: 
\[
K_- = \frac12 \hat{b}^2, \quad K_+ = K_-^\dag, \quad K_3 = \frac12\left( \hat{b}^\dag \hat{b} +\frac12\right).
\]
 The commutation relations 
\[
[K_-,K_+] = 2K_3, \quad [K_3,K_\pm]=  \pm K_\pm
\]
  are used to obtain a closed-form expression for the  unitary evolution  operator: 
\begin{eqnarray}
\label{SU11}
&& e^{-ir(K_- + K_+)} = e^{-iu(r)K_+}e^{-v(r)K_3}e^{-iu(r)K_-}, \nonumber\\
&& \quad r \equiv  2\alpha  \Omega t,\; u = \tanh r, \; v = 2\ln(\cosh r). 
\end{eqnarray}
Applying the right-hand side of Eq. (\ref{SU11}) to the vacuum state in the signal mode results in the standard parametric approximation, which corresponds to the squeezed state \cite{ReviewSq}:
\begin{eqnarray}
\label{Sr}
&& |Sr\rangle \equiv e^{-ir(K_- + K_+)}|0\rangle \\
&&= \sqrt{\mathrm{sech} r} \sum_{n=0}^\infty \binom{2n}{n}^\frac12\left( \frac{-i\tanh r}{2}\right)^n|2n\rangle, \nonumber
\end{eqnarray}
where   $|2n\rangle$ denotes the Fock state with $2n$ photons.  

To compare the above-described parametric approximation with the exact solution, we map the joint state of the pump-signal system onto the invariant subspaces 
$\cH_N \equiv \mathrm{Span}\{|N-n,2n\rangle, 0\le n \le N\}$, where $|N-n,2n\rangle$  is   the Fock state with $N-n$ photons in the pump mode and $2n$ in the signal mode.   
Expanding the  coherent state    over the Fock states $|M\rangle = \frac{(\hat{a}^\dag)^M}{\sqrt{M!}}|0\rangle$, as in Eq. (\ref{CohSt}),  and rearranging the two summations by introducing a new index 
$N \equiv M+n$,   we obtain:
\begin{eqnarray}
\label{full_Sq}
&&|\alpha\rangle|Sr\rangle = e^{-\frac{\alpha^2}{2}}\sum_{M=0}^\infty\frac{\alpha^M}{\sqrt{M!}}|M\rangle|Sr\rangle\nonumber\\
&&=e^{-\frac{\alpha^2}{2}}\sum_{N=0}^\infty\frac{\alpha^N}{\sqrt{N!}}\sum_{n=0}^N \widetilde{\psi}^{(N)}_n |N-n,2n\rangle,
\end{eqnarray}
where the quantum amplitude $\widetilde{\psi}^{(N)}_n$, which serves as a parametric approximation to the exact amplitude  $\psi^{(N)}_n$  from Eqs. (\ref{psit})-(\ref{amplit}) of Section \ref{sec3}, is given by:
\be
\widetilde{\psi}^{(N)}_n  = \left[ (N)_n\binom{2n}{n}\mathrm{sech} r  \right]^\frac12\left( \frac{-i\tanh r}{2\alpha}\right)^n,
\en{psi_p}
where $(N)_n = N(N-1)\ldots (N-n+1)$. 

The parametric approximation to the coefficient ${\gamma}^{(N)}_n$ of the exact solution in Eq. (\ref{psi-state}), for the original quantum Hamiltonian
\[
 \hat{H}_1 = \hbar\Omega(\hat{a}^\dag\hat{b}^2 + \hat{a}\hat{b}^\dag{}^2),
 \]
 can be derived from the relation in Eq. (\ref{amplit}). Noting that in this case $m=1$ and $q=0$ (so that $M=N$)  in Eq. (\ref{beta_1}),  we obtain
\begin{eqnarray}
\label{beta_sq}
&& \beta^{(N)}_n = (N-n)(2n+1)(2n+2),\nonumber\\
&& \prod_{l=0}^{n-1}\beta^{(N)}_l = (N)_n(2n)!.
\end{eqnarray}
From Eqs. (\ref{amplit}) and (\ref{psi_p}), we then obtain the corresponding coefficients in the parametric approximation as functions of the dimensionless time $\tau = \Omega t = {r}/{2\alpha}$:
\be
\widetilde{\gamma}_n =  \sqrt{\mathrm{sech} r}   \frac{\left( \frac{\tanh r}{2\alpha}\right)^n}{n!}.
\en{gam_p}

Note that the parametric approximation $\widetilde{\gamma}_n$ is independent of the subspace dimension parameter $N$. This independence arises because the parametric approximation is valid for $\alpha\gg 1$, meaning that only the subspaces $\cH_N$ with $N$ satisfying Eq.~(\ref{ToREF}) contribute significantly to the result.

Now, let us compare the parametric approximation in Eq. (\ref{gam_p}), which applies for $\alpha\gg 1$, with the exact result in Eq. (\ref{psi-state}) of Section \ref{sec3}. Using the standard Taylor series expansion of the elementary functions in Eq. (\ref{gam_p}), we obtain, up to an error of order $\mathcal{O}(n^2r^4)$:
 \be 
  \widetilde{\gamma}_n = \frac{\tau^n}{n!}\biggl( 1 - \left[\frac{n}{3} +\frac{1}{4}\right]r^2 + \mathcal{O}(n^2r^4)   \biggr).
 \en{gam_ser_p}
On the other hand, expanding Eq. (\ref{psi-state}) up to the same order of error gives the exact expression:
  \be 
  \gamma^{(N)}_n = \frac{\tau^n}{n!}\biggl( 1 - \left[\frac{n}{3} +\frac{1}{4} - \frac{n(n+1)}{4N}\right] \frac{N r^2}{\alpha^2}  
 +  \mathcal{O}(n^2r^4)          \biggr).
  \en{gam_ser_ex}   
  Comparing the expansions in Eqs. (\ref{gam_ser_p}) and (\ref{gam_ser_ex}), we conclude that the parametric approximation has the following relative error (in the significant interval around 
 $\langle N\rangle =\alpha^2$, as discussed above):
  \be
 \frac{ \widetilde{\gamma}_n - \gamma^{(N)}_n}{ \gamma^{(N)}_n} = \mathcal{O}\left( r^2\frac{n+1}{\alpha}\right).
 \en{rel_err}
  Thus, the parametric approximation accurately describes the quantum amplitudes in Eq. (\ref{full_Sq}) for  $n\ll \alpha/r^2= \sqrt{\langle N \rangle}/r^2$.  
  
 According to Eq. (\ref{rel_err}), the parametric approximation fails to accurately approximate the exact amplitudes $\psi^{(N)}_n$ to a given relative error $\mathcal{O}(\epsilon)$ for $n \ge  n_c(\epsilon,r,\alpha)$, where
 \be
 n_c  \sim \frac{\epsilon \alpha}{r^2}.
 \en{GTP8}
   This implies the existence of a maximum value $r_c(\epsilon,\alpha)$ of the squeezing parameter,  for  the parametric approximation   to have    a relative error   $\mathcal{O}(\epsilon)$. The critical value 
 $r_c(\epsilon,\alpha)$ can be estimated by considering the contribution to the norm of the quantum state in Eq. (\ref{full_Sq}) arising from the quantum amplitudes with 
 $n\ge n_c(\epsilon,r,\alpha)$.

A detailed analysis of the applicability conditions of the parametric approximation to the spontaneous down-conversion process,  given its significance for quantum technology, will be considered elsewhere.

  \section{Concluding remarks}
\label{sec5}  
  
 In this work, we have  derived the exact solution for a broad class of nonlinear models describing interacting bosons, with immediate applications to the state evolution problem in quantum optics. Specifically, we have considered a pump mode propagating through a nonlinear optical medium that satisfies the phase-matching conditions, thereby generating signal mode(s) initially in the vacuum state. The solution to the state evolution problem is expressed as a series expansion in powers of the propagation time, where a single function -- polynomial in the signal mode(s) populations -- characterizes the specific quantum model. These results have direct applications to nonlinear models in quantum optics, including the $k$-photon down-conversion process, which plays a crucial role in advancing quantum technology.

Several open problems remain for future research. One key challenge is how to generalize the method to solve the state evolution problem for an arbitrary initial state. Another open question concerns the semiclassical approach commonly used in quantum optics: What is the correct asymptotic limit of the exact solution when the dimension of the invariant subspace (e.g., the average number of photons in the pump mode) approaches infinity? It is hoped that the present approach has laid the foundation for addressing such open problems.
\medskip 
 \section{acknowledgements}
This work was partially supported by the National Council for Scientific and Technological Development (CNPq) of Brazil.  
The author is greatly indebted to an anonymous referee for comments and suggestions. 


\end{document}